\documentclass[aps,prl,a4paper,twocolumn,superscriptaddress,floatfix,showpacs]{revtex4}

\usepackage{graphicx}% Include figure files
\usepackage{dcolumn}% Align table columns on decimal point
\usepackage{bm}% bold math

\begin{document}

\title{Trapping and manipulating neutral atoms with electrostatic fields}
\author{P.~Kr\"uger}
\email[Contact: ]{krueger@physi.uni-heidelberg.de}
\homepage[~\\Website: ]{http://www.atomchip.net}
\author{X.~Luo}
\author{M.~W.~Klein}
\author{K.~Brugger}
\author{A.~Haase}
\author{S.~Wildermuth}
\author{S.~Groth}
\affiliation{Physikalisches Institut, Universit\"at Heidelberg,
69120 Heidelberg, Germany}
\author{I.~Bar-Joseph}
\affiliation{Department of Condensed Matter Physics, The Weizmann
Institute of Science, Rehovot 76100, Israel}
\author{R.~Folman}
\author{J.~Schmiedmayer}
\affiliation{Physikalisches Institut, Universit\"at Heidelberg,
69120 Heidelberg, Germany}

\date{\today}

\begin{abstract}
We report on experiments with cold thermal $^7$Li atoms confined
in combined magnetic and electric potentials. A novel type of
three-dimensional trap was formed by modulating a magnetic guide
using electrostatic fields. We observed atoms trapped in a string
of up to six individual such traps, a controlled transport of an
atomic cloud over a distance of 400$\mu$m, and a dynamic splitting
of a single trap into a double well potential. Applications for
quantum information processing are discussed.
\end{abstract}


\maketitle

The controlled manipulation of cold neutral atoms in potentials
created by substrate mounted microstructures has been progressing
rapidly during the recent years \cite{Fol02}. A variety of atom
optical devices, e.g.\ tightly confining traps and guides
\cite{Rei99-3398,Fol00}, beam splitters \cite{Cas00}, and mirrors
\cite{Hin99-R119} have been realized on such {\em atom chips}.
Issues of current investigations include the behavior of
Bose-Einstein condensates in atom chip potentials \cite{Hae01b},
the coherent quantum dynamics and decoherence processes in traps
and guides near surfaces \cite{Hen03}, and the integration of new
components for the preparation, manipulation, and detection of
atomic quantum states \cite{Hor03}.

In this letter, we present the integration of charge carrying
structures into atom chips. Utilizing electric fields in addition
to the so far exclusively used magnetic fields enhances the
functionality of atom chips by introducing an independent degree
of freedom to the potentials. In our experiments, we have used
electrostatic fields to form a novel type of combined
electro-magnetic trap where the confinement in two dimensions is
magnetic while the movement in the third dimension is controlled
by electric fields. In a string of up to six such traps we could
arbitrarily switch the individual traps on and off, or split a
single trap dynamically into two, thereby demonstrating a new type
of atom chip beam splitter. We employed dynamic electric fields to
transport a cloud of trapped atoms along a magnetic guide in a new
type of atom `motor'.

The interaction of electrostatic fields and neutral but
polarizable atoms (polarizability $\alpha$) is given by the
interaction energy $U_\mathrm{el}$ of the field $E$ and the
induced dipole:
\begin{equation}
U_\mathrm{el}=-\frac{\alpha}{2}E^2 \label{eq:epot}
\end{equation}
Eq.\ \ref{eq:epot} shows that atoms in the ground state are drawn
towards higher electric fields since the interaction of the
induced dipole with the field is always attractive. As a
consequence, the {\em minimum} of a purely electrostatic trapping
potential will be located at a (local) {\em maximum} of $E$.
According to the Earnshaw theorem, however, such a maximum cannot
exist in free space. Any trap utilizing electric fields therefore
has to compensate for the attraction of the trapped atoms towards
the charged objects creating the electric fields \footnote{The
motion of atoms near an attractive $1/r^2$ singularity has been
studied both theoretically \cite{Den97} and experimentally
\cite{Den98}. It has been shown that there are no stable orbits
for the atoms in such a potential.}.

Stabilizing such a trap is possible either by adding (static)
repulsive potentials, e.g. by introducing inhomogeneous magnetic
fields or light fields, or by dynamically changing the electric
fields \cite{Hau92}. For example, it has been suggested to realize
the former by mounting charged structures onto magnetic or optical
atom mirrors \cite{Sch98,Hin99-R119}.

\begin{figure}[h]
    \includegraphics[width= \columnwidth]{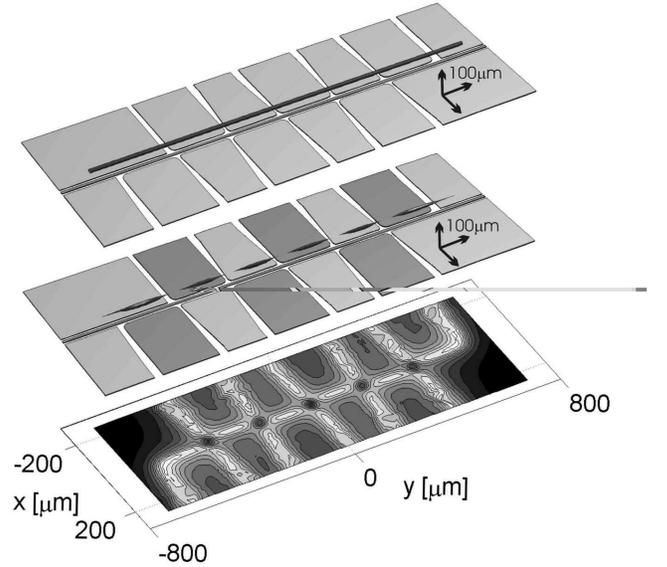}
    \caption{\label{fig:design}Chip design: A regular side guide
    is modulated by applying voltages to the set of six
    electrodes (dark gray in center plot). Equipotential surfaces
    ($E_\mathrm{pot}=k_B \times 475\mu$K) are shown for $^7$Li atoms
    in the $|F=2,m_F=2\rangle$ ground state
    for the side guide (top) and
    a string of six combined electro-magnetic traps (center). A
    contour plot of the electric field strength at the plane parallel to
    the chip at the height of the minimum ($z=50\mu$m) is also depicted
    (bottom). The parameters chosen for these numerical
    calculations were $I_\mathrm{wire,y}=1$A, $B_\mathrm{bias,x}=40$G,
    and $V_\mathrm{elec}=+475$V.}
\end{figure}

In our experiment we combine the electric and magnetic
interactions to realize stable trapping configurations on atom
chips. We produce electric fields by charging a set of electrodes
distributed along a current carrying wire (Fig.\
\ref{fig:design}). The wire produces, together with a homogenous
external bias field, a two-dimensionally confining {\em side
guide} potential \cite{Fol02}. Charging the electrodes results in
an inhomogeneous electric field that is modulated in strength
along the guide. Such a modulation can lead to a full
three-dimensional confinement given by the potential
\begin{equation}
U_\mathrm{mag+el}=\mu_B g_F m_F B - \frac{\alpha}{2}E^2
\label{eq:comb}
\end{equation}
with the magnetic field modulus $B$, the Bohr magneton $\mu_B$,
the Land\'{e} factor $g_F$, and the magnetic quantum number $m_F$
of the trapped atomic state. Fig.\ \ref{fig:design} shows our
design and typical equipotential surfaces of the magnetic side
guide and a string of combined traps that is formed when a
moderate voltage is applied to the six electrodes along the
guiding wire. We have numerically investigated the parameter
ranges needed to form combined electro-magnetic traps with this
geometry and found that a few hundred volts are needed in order to
form potential wells that are deep enough to trap thermal $^7$Li
atoms of $T\sim100\mu$K at a height of $50\mu$m above the chip
surface. If the voltage is too low, the modulation of the guiding
potential becomes weaker, if it is too high, the attractive
electric interaction lowers the magnetic barrier to the chip, and
the atoms can propagate onto the surface. However, the voltage
range where traps are formed is large, and trap depths of up to
$k_B\times 500\mu$K are possible. In this case, the barrier to the
attractive potential near the electrodes is large enough to
inhibit tunnelling to the surface completely.

The two atom chips we have used in our experiments are based on
silicon and sapphire substrates with a thickness of 600$\mu$m. A
patterned gold layer (thickness $\sim4.5\mu$m) was evaporated onto
the substrates by standard lithographic and lift-off techniques.
After the fabrication, a highly reflecting gold mirror is obtained
with the current and charge carrying structures defined by
10--50$\mu$m wide groves in the gold layer providing electrical
isolation \footnote{Both types of substrates tolerated voltage
differences of a few hundred volts over spatial separations of
10$\mu$m. The Si substrate, however, exhibited a slow degradation
at $V_\mathrm{elec}>300$V which was not encountered on the
sapphire substrate at all tested voltages, i.e. up to 400V.}.
\begin{figure}
    \includegraphics[width=\columnwidth]{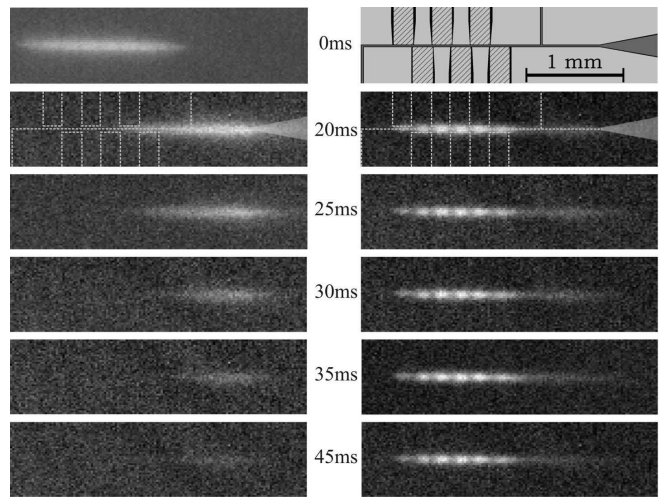}
    \caption{\label{fig:dump}Time series of fluorescence signals in experiments
    demonstrating the trapping of thermal $^7$Li atoms in combined
    electro-magnetic traps: (left) At $t=0$, a cloud of atoms is released
    from a magnetic Z-trap into an open L-shaped guide where the
    guiding wire is broadened at the open end on the right hand side, guiding the atoms
    towards the chip surface where they are lost. (right) If a
    voltage is applied to the electrodes along the guide, atoms
    remain in six individual trap sites. (top right) The actual
    chip design used in these experiments with the different
    current carrying and charged structures. The parameters used
    in this case were $I_\mathrm{wire}=1.6$A, $B_\mathrm{bias}=44$G,
    and $V_\mathrm{elec}=+300$V.}
\end{figure}
In our experiment, we accumulate typically $10^8$ $^7$Li atoms in
a reflection MOT where the chip serves as a mirror. With the
intermediate support of a U-shaped silver wire mounted directly
underneath the chip, the atoms are loaded into a Z-shaped wire
trap on the chip. These magnetic traps and the loading procedure
are described in \cite{Fol00,Cas00}. Fluorescence images are taken
by exposing the atoms to a near resonant light pulse of 100$\mu$s
duration from two counterpropagating light beams that enter the
vacuum chamber parallel to the chip surface. This configuration
suppresses all artifacts on the images stemming from features on
the chip surface. Even if electric fields are present at the time
of the imaging light pulse, the resonance shift due to the
differential Stark effect is much smaller than the line width of
the atomic transition (at a typical field of 10kV/cm,
$\Delta\nu_\mathrm{Stark}=450$kHz $\ll \Gamma/2\pi=5.8$MHz
\cite{Win92}). Therefore, the modulation of the fluorescence
signal can only be attributed to an actual modulation in the
atomic density.

After loading, the atoms hover 50--100$\mu$m above the chip
surface in an elongated trap of approximately 1mm in length. In
the next step, the trapping potential is opened on one side by
switching the current so that it now runs along an L-shaped path.
If no voltage is applied to the electrodes along the central piece
of the Z-shaped wire, all atoms are lost within approximately 50ms
after they are released to the guide (Fig.\ \ref{fig:dump} left).
If, however, at the time of the opening of the guide the voltage
on the electrodes is switched on non-adiabatically (within
5$\mu$s) to 250--350V, a sizeable fraction of the cloud, i.e.
those (sufficiently cold) atoms that are located above the
electrodes where the potential minima appear, remains trapped
(Fig.\ \ref{fig:dump} right). A comparison of the two experiments
described above clearly shows the trapping effect of the electric
field introduced by the high voltage. All six electrodes could be
switched separately so that an arbitrary subset of traps could be
operated.

The experimental results agree quantitatively with the numerical
predictions presented above regarding size and position of the
traps, while the exact voltages leading to a maximal number of
trapped atoms differ by up to 50\%. We attribute this deviation to
the fact that the calculations were idealized, i.e. the dielectric
properties of the involved materials were neglected and only the
central region of the chip was taken into account. The anticipated
behavior of weaker modulation in the atom density for lower
electric fields was observed as well as the loss of atoms to the
chip surface at voltages that were too high. The measured
lifetimes of the combined electro-magnetic traps are in agreement
with those of the purely magnetic Z-trap.

In order to demonstrate some of the new capabilities of the atom
chip introduced by the integration of electric fields, we carried
out two further experiments. In the first experiment we showed the
controlled transfer of an atom cloud along a magnetic guide by
means of electric fields (see \cite{Rei99-3398} for a purely
magnetic transfer).
\begin{figure}
    \includegraphics[width=\columnwidth]{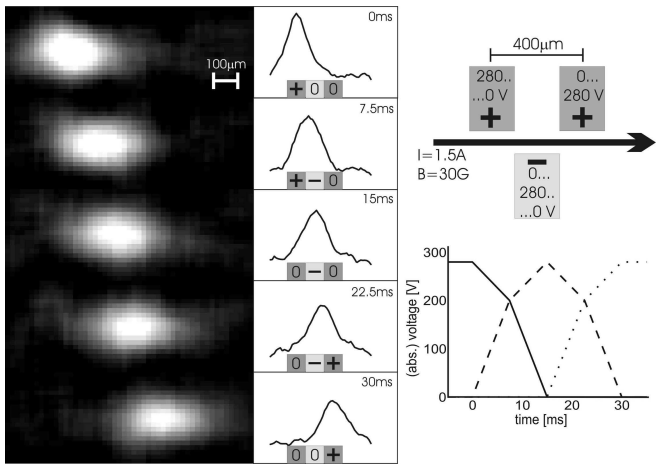}
    \caption{\label{fig:motor}(left) An atom cloud confined in a single
    combined electro-magnetic trap ($I_\mathrm{wire}=1.5$A,
    $B_\mathrm{bias}=30$G, and $V_\mathrm{elec}=280$V) is smoothly
    transported along the magnetic side guide. (center) Longitudinal profile
    of the atom distribution as the cloud is transferred over a
    distance of 400$\mu$m during 30ms. The voltage on the
    different electrodes at the different phases of the transfer
    are depicted schematically. (right) The timing of the voltage
    ramps was chosen to allow a smooth transfer from one trapping
    site to another while maintaining a nearly constant trap depth of
    50--60$\mu$K$\times k_B$.}
\end{figure}
The geometry of the chip structures necessary for this `motor'
experiment and the timing sequence is illustrated in Fig.\
\ref{fig:motor}. A single trap was loaded initially by repeating
the above described procedure, but only switching on the voltage
on one of the electrodes. After a short trapping time, a voltage
of reversed polarity was ramped up on a neighboring electrode on
the opposite side of the guiding wire while the voltage on the
first electrode was ramped down to zero. At the intermediate stage
where the voltages on both electrodes are on, the electric field
strength along the guide is maximal in the middle between the two
electrode centers. The different polarity is essential for a
smooth, barrier-free transfer of the cloud. To maintain a nearly
constant trap depth throughout the transfer, the voltage ramps
were run in two steps: During the first 7.5ms the voltage on the
first electrode was ramped from $+280$V to $+200$V, while the
voltage on the second electrode was ramped from 0 to $-200$V.
Within the following 7.5ms, the first voltage was ramped down to
0, while the second voltage was ramped up completely to $-280$V.
This process was subsequently replicated between the second
electrode and a third electrode, again located on the opposite
side of the guiding wire. After the complete sequence of 30ms
duration, the whole atomic cloud had been transported over a
distance of 400$\mu$m. The measured loss of atoms during the
transfer is equal to that observed in a static trap, proving that
losses induced by the transfer process are negligible with respect
to the trap lifetime.

\begin{figure}[b]
    \includegraphics[width=\columnwidth]{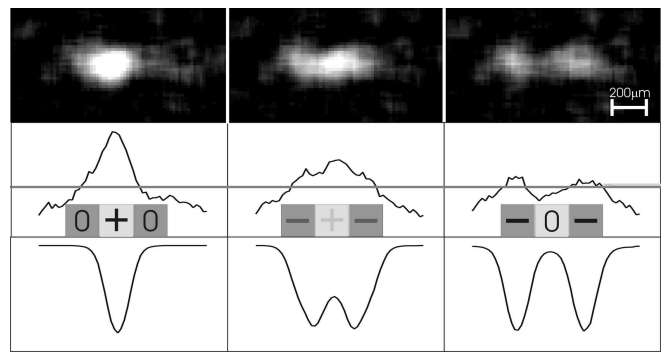}
    \caption{\label{fig:splitting}Dynamic splitting of an atom
    cloud. A single trap is gradually (within 15ms) turned into a
    double well potential by ramping the voltage on one electrode
    down while two neighboring electrodes are slowly charged.
    (top) Fluorescence images of the atoms during different
    stages of the experiment. The inserts schematically show the polarities and
    voltages (darker shading corresponds to higher voltage) during the
    different phases of the experiment. (center) The density profile clearly
    shows the gradual splitting of a single cloud into two.
    (bottom) The corresponding potentials along the axis of free
    motion in the magnetic guide.}
\end{figure}

In the second experiment we observed a dynamic splitting and
recombination of an atom cloud into two separate clouds. Again,
the experiment was initiated by loading a single trap. While the
voltage on the first electrode is ramped down to 0, voltages of
equal magnitude but different polarity ($-200$V) are gradually
(within 15ms) ramped up on the two neighboring electrodes on the
opposite side of the guiding wire in a fashion similar to the one
used for the `motor'. The results of this experiment and the
calculated potentials during the different stages of the
experiment are shown in Fig.\ \ref{fig:splitting}. For the
recombination of the two clouds, the process is reversed. Also in
these experiments, no additional loss with respect to the
lifetimes in the static traps could be detected.

Finally, we give examples of important future applications of
electric fields on atom chips. The first relies on the fact that
the magnetic part of the combined electro-magnetic potential (eq.\
\ref{eq:comb}) depends on the magnetic quantum number $m_F$ while
the electric part does not. This gives rise to possibilities of
state-selective operations which are essential for quantum
information processing (QIP). In \cite{Cal00a} it has been shown
that a two-qubit phase gate based on controlled collisions can be
realized by storing two atoms in the two minima of a double well
potential in which the separating barrier can be conditionally
removed for one qubit state but not for the other. If the qubit is
encoded in atomic hyperfine states with different $m_F$, a
combination of electric and magnetic potentials will allow such a
configuration: The central barrier in a magnetic double well
potential is of different height for the different $m_F$, and can
therefore be removed state-selectively by an electric field since
the electric potential does not depend on $m_F$ (eq.\
\ref{eq:epot}).

The second application, again of significance especially for
prospects of QIP on atom chips, actually exploits the differential
Stark effect which does not play a role in the current
experiments: Atom chips offer the possibility to design trapping
structures with a sub-micron spatial resolution. An array of traps
with inter-trap distances of the same order will be desirable in
scaled QIP-experiments. However, individual addressability of the
trapping sites by laser light beams would be limited by the
achievable beam waist. This problem could be overcome by
introducing switchable electric fields at each trap site that
would be used to shift the atoms in and out of resonance with the
light frequency. In contrast to the current experiments where the
goal was to trap atoms which requires large {\em gradients} in
$U_\mathrm{el}$ at relatively low field strengths, the charged
structures used here should produce strong (for sufficiently large
detuning) and more locally homogeneous fields in order to keep the
potential shape unaltered.

A third application of potentials derived from electric fields is
to independently alter the phase evolution of atoms (qubits) in
magnetic traps and guides in a controlled way. This is of
particular interest for various types of atom chip based
interferometers \cite{Fol02}.

To conclude, we have demonstrated for the first time the trapping
and manipulation of atoms using electrostatic fields. This adds a
new degree of freedom to quantum optics on atom chips and paves
the way to a variety of novel experiments and applications.

This work was supported by the European Union, contract numbers
IST-2001-38863 (ACQP) and HPRI-CT-1999-00069 (LSF) and the
Deutsche Forschungsgemeinschaft, Schwerpunktprogramm
`Quanteninformationsverarbeitung'.

\end{document}